\documentclass[twocolumn,floatfix,eqsecnum,aps,prb]{revtex4-2}
\usepackage{graphicx}
\usepackage{amsmath}
\usepackage{amssymb}
\usepackage{bm}
\usepackage{hyperref}

\usepackage{color}

\begin{document}

\title{Shot noise from which-path detection in a chiral Majorana interferometer}
\author{C. W. J. Beenakker}
\affiliation{Instituut-Lorentz, Universiteit Leiden, P.O. Box 9506, 2300 RA Leiden, The Netherlands}

\date{May 2025}

\begin{abstract}
We calculate the full counting statistics of charge transfer in a chiral Majorana interferometer --- a setup where a Dirac mode (an electron-hole mode) is split into two Majorana modes that encircle a number of $h/2e$ vortices in a topological superconductor. Without any coupling to the environment it is known that the low-energy charge transfer is deterministic: An electron is transferred either as an electron or as a hole, dependent on the parity of the vortex number. We show that a stochastic contribution appears if which-path information leaks into the environment, producing the shot noise of random $2e$ charge transfers with binomial statistics. The Fano factor (dimensionless ratio of shot noise power and conductance) increases without bound as the which-path detection probability tends to unity.
\end{abstract}
\maketitle

\section{Introduction}

\begin{figure}[tb]
\centerline{\includegraphics[width=1\linewidth]{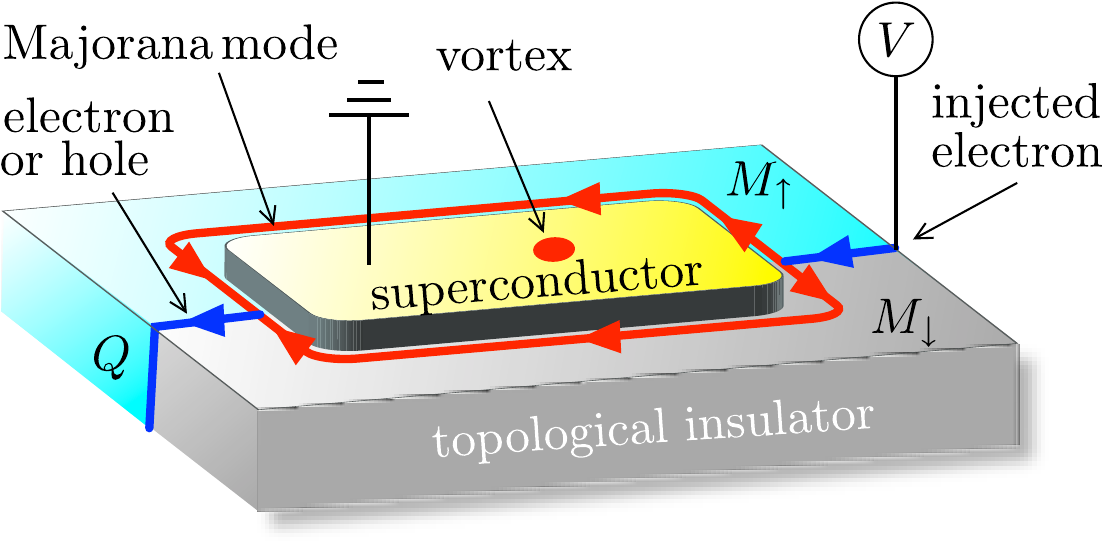}}
\caption{Layout of the $\mathbb{Z}_2$ interferometer \cite{Fu09,Akh09}. A three-dimensional topological insulator is covered by magnetic insulators with opposite polarization ($M_\downarrow$ and $M_\uparrow$) and by a superconductor. A pair of chiral Majorana modes flow along the edge between superconductor and magnet. A bias voltage $V$ injects an electron into a superposition of the two Majorana modes, which then recombine either as an electron or as a hole after having encircled a certain number of $h/2e$ vortices. The transferred charge $Q$ depends on the $\mathbb{Z}_2$-valued parity of the number of vortices.
}
\label{fig_layout}
\end{figure}

Because a Majorana fermion has a real wave function, only phase shifts $0$ or $\pi$ are allowed in a scattering process \cite{Alt97,Bee16}. This restriction is at the basis of the $\mathbb{Z}_2$ interferometer in a topological superconductor \cite{Fu09,Akh09,Law09,Str11,Li12,Par14,Str15,Sha21}: A chiral Dirac mode splits into a pair of chiral Majorana modes, which recombine after having encircled a number $N_v$ of $h/2e$ vortices (see Fig.\ \ref{fig_layout}). Depending on the $\mathbb{Z}_2$-valued parity of $N_v$, a low-energy electron (charge $+e$) is transferred through the interferometer either as a $+e$ electron or as a $-e$ hole. (In the latter case the missing $2e$ charge is absorbed by the superconducting condensate.) This charge transfer is noiseless, fully deterministic.

In an electronic Aharonov-Bohm interferometer the oscillatory magnetic field dependence of transport properties is damped by decoherence due to coupling to the environment. The mechanism is which-path detection \cite{Buk98,Spr00}: If the pathway taken by the electron through the interferometer leaves a trace in the environment, no quantum interference of different pathways can occur. Here we wish to examine this effect in the $\mathbb{Z}_2$ interferometer. We find that which-path detection adds a stochastic contribution to the charge transfer statistics, producing shot noise as a signature of decoherence in Majorana interferometry.

The present study is an application to a topological superconductor of a general method which I recently developed with Jin-Fu Chen \cite{Bee25}. A chiral interferometer is modeled by a monitored quantum channel \cite{NielsenChuang}, in which unitary propagation alternates with weak measurements of the occupation number. It was shown in Ref.\ \onlinecite{Bee25} for a quantum Hall interferometer that the generalized Levitov-Lesovik formula \cite{Lev93,Lev96} following from this quantum-information based approach preserves the binomial form of the charge transfer probability expected for Fermi statistics \cite{Bla00}, thereby removing a shortcoming of the dephasing-probe model of decoherence \cite{Mar04,Chu05,Pil06,For07,Hel15}. 

Here we find as well that the full counting statistics is binomial, with random charge transfers of size $2e$ rather than $e$, associated with the presence of a superconducting condensate. The Fano factor $F$, the dimensionless ratio of shot noise power and conductance \cite{Bla00}, of the Majorana interferometer is anomalous, very different from the quantum Hall effect: We find that $F$ increases monotonically from zero to arbitrarily large values as the which-path detection becomes more and more effective. 

\section{Full counting statistics of the $\pmb{\mathbb{Z}_2}$ interferometer}

The central object for the full counting statistics is the cumulant generating function $C(\xi)$ of the number of electron charges transferred in a time $t_{\rm counting}$, in the limit $t_{\rm counting}\rightarrow\infty$. We assume elastic scattering, so we can consider separately the contribution of each energy $E$ to the cumulants,
\begin{equation}
C(\xi)=\sum_E \ln F(\xi,E),\;\;F(\xi,E)=\langle e^{\xi Q(E)}\rangle,\label{CxiFxi}
\end{equation}
with $F(\xi,E)$ the moment generating function of the charge $Q(E)$ transferred at energy $E$. Charge is measured in units of the electron charge $e$ and energy is discretized in units of $\delta E=h/t_{\rm counting}$. At zero temperature and a bias voltage $V>0$, energies in the range $0<E<eV$ contribute. We will now focus on a single $E>0$, and then at the end sum over all energies.

\begin{figure}[tb]
\centerline{\includegraphics[width=0.7\linewidth]{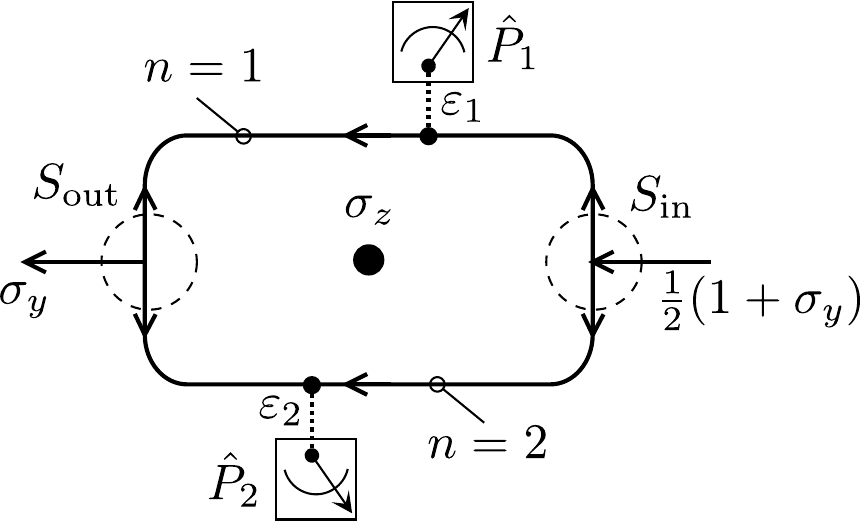}}
\caption{Scattering geometry corresponding to Fig.\ \ref{fig_layout}. The operators $\hat{P}_1$ and $\hat{P}_2$ are weak measurements of the occupation number of chiral Majorana modes $n=1,2$. The vortex (black dot) inserts a $\pi$ phase shift between the two modes (Pauli matrix $\sigma_z$). At the entrance the matrix $(1+\sigma_y)/2$ projects onto an incoming electron, at the exit the matrix $\sigma_y$ measures the transferred charge. All of this is in the Majorana basis, in the electron-hole basis the role of  $\sigma_y$ and $\sigma_z$ is interchanged.
}
\label{fig_layout2}
\end{figure}

Following the general method of Ref. \onlinecite{Bee25}, we consider a weak measurement of the occupancy of the Majorana mode in each of the arms of the interferometer: mode number $n=1$ in the upper arm, mode number $n=2$ in the lower arm, see Fig.\ \ref{fig_layout2}. The measurement of mode $n$ interpolates with a weight factor $\varepsilon_n$ between the identity $\hat{I}$ and a projection onto a filled ($\hat{P}_{+,n}$) or empty ($\hat{P}_{-,n}$) mode:
\begin{equation}
\begin{split}
&\hat{P}_{+,n}=\delta_n\hat{I} +\varepsilon_n a_n^\dagger a_n^{\vphantom{\dagger}},\;\;\hat{P}_{-,n}=\delta_n\hat{I}+\varepsilon_n a_n^{\vphantom{\dagger}}a_n^\dagger,\\
&\delta_n=\tfrac{1}{2}(\sqrt{2-\varepsilon_n^2}-\varepsilon_n),\;\;0\leq \varepsilon_n\leq 1.
\end{split}\label{projectors}
\end{equation}
The operators $a_n^{\vphantom{\dagger}},a_n^\dagger$ are Majorana fermion operators at $E>0$, with anticommutation relations \cite{note1}
\begin{equation}
\{a^\dagger_n,a_m\}=\delta_{nm},\;\;\{a_n,a_m\}=0.\label{anticommutator}
\end{equation}
The coefficients $\varepsilon_n,\delta_n$ are chosen such that  
\begin{equation}
\hat{P}_{+,n}^2+\hat{P}_{-,n}^2=\hat{I}.\label{Psumrule}
\end{equation}
The which-path measurement \eqref{projectors} is preceded and followed by unitary propagation, with scattering operators $\hat{S}_{\rm in}$ and $\hat{S}_{\rm out}$. 

The incoming modes are in thermal equilibrium at inverse temperature $\beta$, 
\begin{equation}
\hat\rho_{\rm in}=Z^{-1}e^{-\beta a^\dagger Ha},\;\;Z=\operatorname{Tr}e^{-\beta a^\dagger Ha},
\end{equation}
with single-particle Hamiltonian $H$. (We collect the operators $a_n^{\vphantom{\dagger}},a_n^\dagger$ in vectors $a,a^\dagger$, so that $a^\dagger Ha=\sum_{n,m}a^\dagger_n H_{nm}a_m$.) The outgoing modes have the density matrix of a monitored quantum channel \cite{Bee25},
\begin{equation}
\begin{split}
&\hat{\rho}_{\rm out}=\sum_{s_1,s_2=\pm}\hat{\cal K}_{s_1,s_2}\hat{\rho}_{\rm in}\hat{\cal K}_{s_1,s_2}^\dagger,\\
&\hat{\cal K}_{s_1,s_2}=\hat{S}_{\rm out}\hat{P}_{s_1,1}\hat{P}_{s_2,2}\hat{S}_{\rm in}.
\end{split}\label{channel}
\end{equation} 
The Kraus operators $\hat{\cal K}_\pm$ satisfy the sum rule
\begin{equation}
\sum_{s_1,s_2=\pm}\hat{\cal K}^\dagger_{s_1,s_2}\hat{\cal K}^{\vphantom{\dagger}}_{s_1,s_2}=\hat{I},
\end{equation}
in view of Eq.\ \eqref{Psumrule} and the unitarity of $\hat{S}$. This ensures that $\operatorname{Tr}\hat{\rho}_{\rm out}=\operatorname{Tr}\hat{\rho}_{\rm in}=1$; the quantum channel \eqref{channel} is a completely-positive trace-preserving map \cite{NielsenChuang}.

The elastic scattering operator $\hat{S}$ at energy $E$ is the exponent of a quadratic form in the fermionic operators, 
\begin{equation}
\hat{S}=e^{ia^\dagger L(E)a},
\end{equation}
with $L(E)$ a Hermitian $2\times 2$ matrix. The unitary matrix $S(E)=e^{iL(E)}$ is the single-particle scattering matrix. The two scattering operators $\hat{S}_{\rm in}$ and $\hat{S}_{\rm out}$ are thus represented by a pair of unitary $2\times 2$ matrices $S_{\rm in}$ and $S_{\rm out}$.

The charge operator $\hat{Q}$ in the Majorana basis is 
\begin{equation}
\hat{Q}=i\bigl(a_2^\dagger a_1^{\vphantom{\dagger}}-a_1^\dagger a_2^{\vphantom{\dagger}}\bigr)=a^\dagger \sigma_ya.
\end{equation}
(The Pauli matrix $\sigma_y$ would be $\sigma_z$ in the electron-hole basis.) The moment generating function $F(\xi,E)$ of the transferred charge at energy $E$ is given by
\begin{align}
&F(\xi,E)=\operatorname{Tr}\hat{\rho}_{\rm out}e^{\xi\hat{Q}}\nonumber\\
&\qquad=Z^{-1}\sum_{s_1,s_2=\pm}\hat{\cal K}_{s_1,s_2}e^{-\beta a^\dagger Ha}\hat{\cal K}_{s_1,s_2}^\dagger e^{\xi a^\dagger\sigma_y a}.
\end{align}

At this stage it is helpful to assume $\varepsilon_n\neq 1$. (We can later reach $\varepsilon_n=1$ by taking the limit.) For $0\leq \varepsilon_n<1$ the projectors $\hat{P}_{\pm,n}$ have a Gaussian representation \cite{Bee25},
\begin{equation}
\begin{split}
&\hat{P}_{\pm,n}=c_{\pm,n} e^{\pm\gamma_n a_n^\dagger a_n^{\vphantom{\dagger}}},\;\;\gamma_n=\ln(1+\varepsilon_n/\delta_n),\\
&c_{+,n}=\delta_n,\;\;c_{-,n}=\varepsilon_n+\delta_n.
\end{split}
\end{equation}

Traces of products of Gaussian operators can be evaluated by means of Klich's trace-determinant relation \cite{Kli03},
\begin{align}
&\operatorname{Tr}e^{a^\dagger A_1a}e^{a^\dagger A_2a}\cdots e^{a^\dagger A_pa}=\operatorname{Det}\bigl(1+e^{A_1} e^{A_2}\cdots e^{A_p}\bigr),\label{Klichformula}\\
&\Rightarrow F(\xi,E)=\sum_{s_1,s_2=\pm}c_{s_1}^2c_{s_2}^2\operatorname{Det}(1+e^{-\beta H})^{-1}\nonumber\\
&\qquad\qquad\qquad\times\operatorname{Det}\bigl(1+ {\cal S}_{s_1,s_2}e^{-\beta H}{\cal S}_{s_1,s_2}^\dagger e^{\xi\sigma_y}\bigr)\nonumber\\
&\qquad=\sum_{s_1,s_2=\pm}c_{s_1}^2c_{s_2}^2\operatorname{Det}\bigl(1+{\cal N}_{\rm in}[{\cal S}_{s_1,s_2}^\dagger e^{\xi \sigma_y}{\cal S}_{s_1,s_2}-1]\bigr).\label{Fxideterminant}
\end{align}
We have defined
\begin{equation}
\begin{split}
&{\cal S}_{s_1,s_2}=S_{\rm out}e^{s_1\gamma_1 |1\rangle\langle 1|}e^{s_2\gamma_2 |2\rangle\langle 2|}S_{\rm in},\\
&{\cal N}_{\rm in}=(1+e^{\beta H})^{-1}.
\end{split}
\end{equation}

In the zero-temperature limit, and for positive bias voltage, only electrons are injected, so that ${\cal N}_{\rm in}$ is a projector,
\begin{equation}
{\cal N}_{\rm in}=\tfrac{1}{2}(1+\sigma_y),
\end{equation}
within the energy range $0<E<eV$.

Eq.\ \eqref{Fxideterminant} generalizes the Levitov-Lesovik formula for full counting statistics \cite{Lev93,Lev96}, to include the effects of weak measurements \cite{Bee25}.

\section{Binomial charge transfer statistics}

The key difference between the quantum Hall interferometer studied in Ref.\ \onlinecite{Bee25} and the Majorana interferometer considered here is the constraint of particle-hole symmetry, which requires that
\begin{equation}
S(-E)=S^\ast(E).
\end{equation}
Both $S_{\rm in}$ and $S_{\rm out}$ contain contributions from the junction where the two Majorana modes combine into a Dirac mode (an electron-hole mode). We represent each junction by a point scatterer, of dimension small compared to $\hbar v_{\rm F}/eV$, with $v_{\rm F}$ the Fermi velocity. The scattering matrix of the junction may then be evaluated at the Fermi level, $E=0$, when it is an element $e^{i\alpha\sigma_y}$ of ${\rm SO}(2)$. 

Propagation along the interferometer does not couple the Majorana modes, the relative phase shift at energy $E$ is $e^{ik\delta L\sigma_z}$, with $k=\tfrac{1}{2}E/\hbar v_{\rm F}$ and $\delta L=L_1-L_2$ the path length difference along the two arms. Each $h/2e$ vortex in addition contributes a $\pi$ relative phase shift, so a factor $\sigma_z^{N_v}$ for $N_v$ vortices. All together we have
\begin{equation}
\begin{split}
&S_{\rm in}=\sigma_z^{N_v} e^{ik\delta L\sigma_z}e^{i\alpha_{\rm in}\sigma_y},\\
&S_{\rm out}=e^{i\alpha_{\rm out}\sigma_y}.
\end{split}
\end{equation}
Since the phase shifts $\propto\sigma_z$ commute with the projective measurements, we may associate them either with $S_{\rm in}$ or $S_{\rm out}$ --- their order relative to the measurements does not matter.

Evaluation of the determinant \eqref{Fxideterminant} gives the result
\begin{align}
&F(\xi,E)=\cosh\xi+s_v(1-p_1)(1-p_2)\cos(2k\delta L)\sinh\xi ,
\end{align}
where $s_v=(-1)^{N_v}$ is the vortex parity and $p_n=\varepsilon_n^2$ is the probability of which-path detection in mode $n$. Notice that the dependence on $p_2$ drops out once $p_1=1$, which is as it should be: Since the injected electron passes through the interferometer either in Majorana mode 1 or in Majorana mode 2, once the occupation of mode 1 is measured, it no longer matters whether the occupation of mode 2 is measured or not.

A final simplification applies at low voltages $V\ll \hbar v_{\rm F}/e\delta L$. We can then replace $\cos (2k\delta L)\approx 1$ in the energy range $0<E<eV$, and the cumulant generating function takes the form
\begin{subequations}
\label{Cbinomial}
\begin{align}
C(\xi)={}&N_{\rm in}\ln\bigl[\cosh\xi+s_v(1-p_1)(1-p_2)\sinh\xi\bigr],
\nonumber\\
={}&N_{\rm in}\bigl(s_v\xi+\ln[1-{\cal T}+{\cal T}e^{-2s_v\xi}]\bigr),\\
{\cal T}={}&\tfrac{1}{2}(p_1+p_2-p_1p_2).
\end{align}
\end{subequations}
The number 
\begin{equation}
N_{\rm in}=eV/\delta E=eVt_{\rm counting}/h
\end{equation}
is the number of electrons injected in the interferometer during the counting time $t_{\rm counting}$, in the large-time limit when the discreteness of $N_{\rm in}$ can be ignored \cite{Has08}.

Eq.\ \eqref{Cbinomial} combines a deterministic transfer of $s_v$ electron charges and a stochastic transfer of $-2s_v$ charges with probability ${\cal T}\in(0,1/2)$. The corresponding probability distribution function of the transferred charge $Q$ is binomial,
\begin{subequations}
\begin{align}
&P(Q)=\binom{N_{\rm in}}{\tfrac{1}{2}(N_{\rm in}-s_v Q)} {\cal T}^{\tfrac{1}{2}(N_{\rm in}-s_v Q)}(1-{\cal T})^{\tfrac{1}{2}(N_{\rm in}+s_v Q)},\\
&Q\in\{-N_{\rm in},-N_{\rm in}+2,\ldots N_{\rm in}-2,N_{\rm in}\}.
\end{align}
\end{subequations}

In the full detection limit $p_1\rightarrow 1$ or $p_2\rightarrow 1$, when ${\cal T}\rightarrow 1/2$, this becomes independent of the vortex number parity $s_v$,
\begin{equation}
P(Q)=2^{-N_{\rm in}}\binom{N_{\rm in}}{\tfrac{1}{2}(N_{\rm in}-Q)},\;\;\text{if}\;\;{\cal T}=\tfrac{1}{2}.
\end{equation}

\begin{figure}[tb]
\centerline{\includegraphics[width=0.7\linewidth]{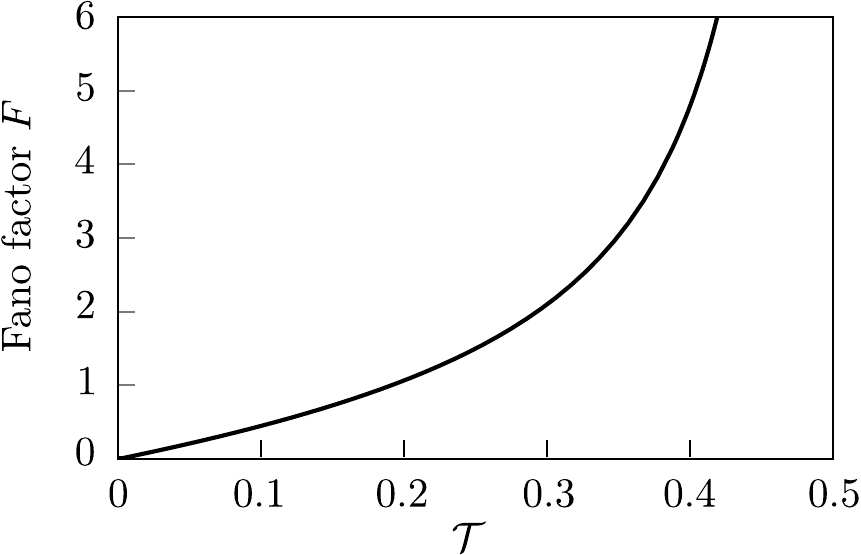}}
\caption{Fano factor as a function of ${\cal T}=\tfrac{1}{2}(p_1+p_2-p_1p_2)$, computed from Eq.\ \eqref{Fano}. This dimensionless ratio of shot noise power and conductance increases without bounds as the which-path detection probability tends towards unity (as ${\cal T}\rightarrow 1/2$).
}
\label{fig_Fano}
\end{figure}

The mean and variance of the transferred charge are
\begin{align}
&\overline{Q}=s_v N_{\rm in}(1-2{\cal T})=s_v N_{\rm in}(1-p_1)(1-p_2),\\
&{\rm Var}\, Q=4N_{\rm in}{\cal T}(1-{\cal T})=N_{\rm in}\bigl(1-(1-p_1)^2(1-p_2)^2\bigr).
\end{align}
In a transport measurement one can access these as the electrical conductance and shot noise power \cite{Bla00}. Their dimensionless ratio, the Fano factor
\begin{equation}
F=\frac{{\rm Var}\,Q}{|\overline{Q}|}=\frac{{\cal T}(1-{\cal T})}{1-2{\cal T}},\label{Fano}
\end{equation}
diverges as ${\cal T}\rightarrow 1/2$ (see Fig.\ \ref{fig_Fano}).

\section{Conclusion}

In summary, we have investigated the effect on the $\mathbb{Z}_2$ interferometer \cite{Fu09,Akh09} of a weak measurement, capable of partially or fully distinguishing whether a Majorana fermion propagates through the upper or lower arm of the interferometer. Such a measurement fundamentally alters the interference by introducing which-path information, thereby degrading the coherence between the two paths.

We find that a which-path measurement adds a stochastic contribution with binomial statistics to the deterministic charge transfer. This contribution is measurable as a shot noise of the electrical current passed through the interferometer in response to a bias voltage. We note that shot noise of an unpaired Majorana mode is known to have a \textit{trinomial} statistics \cite{Str15,Gne15}, distinct from the binomial form found here.

We obtained simple expressions for the full counting statistics, dependent only on the probability of the which-path detection, by working with a high-level description of the measurement as a Gaussian quantum channel \cite{Bee25}. In the present context one source of which-path information is provided by the charge fluctuations of a chiral Majorana mode, coupled to the electromagnetic environment. One would need a microscopic model of this coupling, to find out how effective the which-path measurement is in a realistic geometry. We hope that the striking effect of the coupling presented here, a divergent Fano factor, will motivate such a modeling.

\acknowledgments

This work was supported by the Netherlands Organisation for Scientific Research (NWO/OCW), as part of Quantum Limits (project number {\sc summit}.1.1016). I have benefited from discussions with A. R. Akhmerov and F. Hassler.

\end{document}